\documentclass[11pt]{article}
\usepackage{ben,epsfig}

\def\be{\begin{equation}}
\def\ee{\end{equation}}
\def\bea{\begin{eqnarray}}
\def\eea{\end{eqnarray}}

\begin{document}
\vspace*{4cm}
\title{Simulating clusters of galaxies: a brief history of ``N'' and overmerging}

\author{\bf Ben Moore}

\address{
Department of Physics, Durham University, South
Road, Durham, UK} 

\maketitle\abstracts{
{\bf 
I review four decades of numerical simulations of galaxy clusters focussing on
the attempts to resolve their internal structure. Overmerging describes the
numerical or physical disruption of dark matter halos within dense
environments. This problem was inherent in simulations prior to 1998 but can
be completely overcome with current algorithms and hardware.  We can now
resolve many thousands of subhalos within clusters and we may have converged
on their inner structure, allowing several new tests of the hierarchical
structure formation model and the nature of dark matter.
}
}

\section{Introduction}

The notion that a galaxy cluster forms via gravitational instability within an
expanding universe was first discussed by van Albada (1961). The first
numerical studies of cluster stability with 10-100 particles were carried out
by von Hoerner (1960) and Aarseth (1963).  An investigation of the
virialisation process using several hundred particles was carried out by Henon
(1964) and Peebles (1970). The assumption being that the ``galaxies'' had
already collapsed prior to the turn around of the cluster.  From these first
simulations it was clear that one could form an equilibrium cluster of
particles with global properties not too unlike the theorists ``ideal'' Coma
cluster.

White (1976) followed the evolution of 700 unequal mass particles
expanding with the Hubble flow. In this case it was possible to follow
the collapse of groups of particles prior to the formation of the
cluster. However, these clumps of bound particles were erased during the
collapse and virialisation of the final system.  This work may have been the
key motivation behind the seminal paper ``Core condensation within heavy
halos...'' (White \& Rees 1978). This paper argued that cooling and settling of
baryons at the centres of dark matter halos was essential to the formation and
survival of galaxies in dense environments.  Although a valid description of
galaxy formation, it is now apparent that this process is not 
necessary in order to resolve substructure within hierarchical models.

\

\psfig{figure=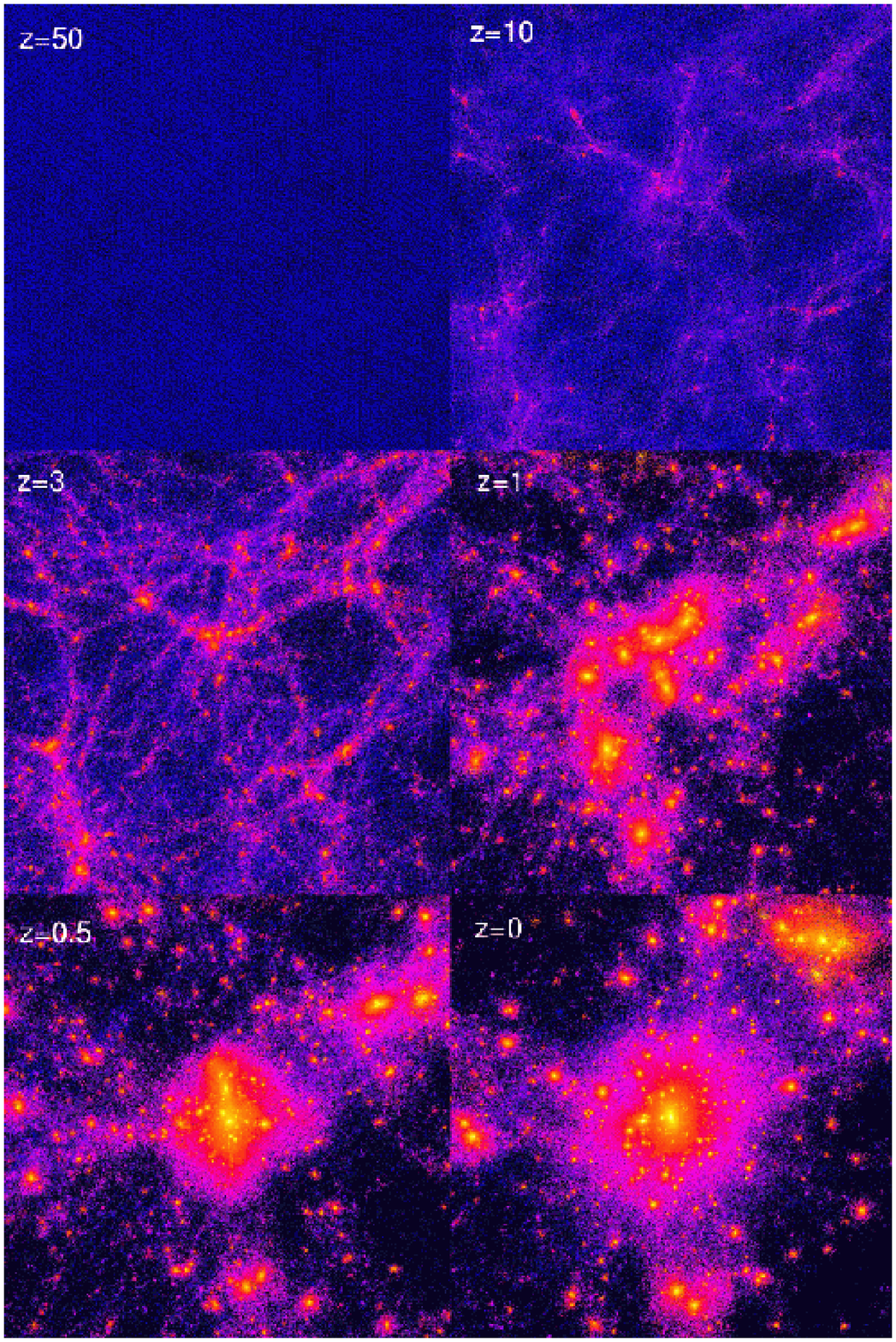,height=7.in}
%\centerline{\epsfysize=8.0truein \epsfbox{pic1_times.ps}}

\

\noindent{\small {\bf Figure 1} \ \ 
The hierarchical evolution of a galaxy cluster in a universe dominated by cold
dark matter.  Small fluctuations in the mass distribution are present but
barely visible at early epochs.  These grow by gravitational instability,
merging and accretion of mass, eventually collapsing into virialised
quasi-spherical dark matter halos. This plot shows a time sequence of 6 frames
of a region of the universe that evolves into a cluster of galaxies.  The
colours represent the local density of dark matter plotted using a logarithmic
colour scale. Linear over-densities are darker blue, whereas the non-linear
collapsed regions attain over-densities of a million times the mean background
density and are plotted as yellow/white.  Each box is 10 Mpc on a side and the
final cluster virial radius is 2 Mpc.
}

\

In the 1980's more detailed cosmological models were developed in which
the mass of the universe was
dominated by collisionless particles. The linear power spectrum of density
fluctuations was determined by the growth of fluctuations in the expanding
universe and the temperature of the dark matter particle. The cold dark matter 
(CDM) model was born (Peebles 1983).

N-body simulations of cluster formation were used to constrain the hot dark
matter model (White etal 1984) and most subsequent work has focussed on the
cold dark matter model (CDM), the currently favoured scenario for structure
formation in the universe. Warm dark matter has not received a great deal of
attention in the literature - primarily because this model could not naturally
produce a closure density of dark matter.  Attention has now shifted from
standard CDM models, which fail for a variety of reasons, to a Lambda
dominated CDM universe. Since the universe is clearly not as simple as one
might wish, alternative models such as warm or self interacting dark matter
may become more popular. Perhaps the widespread acceptance of a positive
Lambda term stems mainly from the fact that the most hierarchical model fails
without its inclusion. Although a hierarchical universe can explain many
fundamental observations, from the microwave background to clustering at high
redshifts, we still know next to nothing about the physical nature of the dark
matter.

Many of the numerical simulations during the 1980's focussed on a comparison
between models and observations of large scale structure. Studies of
individual halo properties were taken from large cosmological volume
simulations. For example, Frenk etal (1985) and Quinn etal (1986) analysed the
structure of dark matter halos that contained just a few thousand particles
per object.  Code development, ``volume renormalisation'' and faster computers
lead to the ability to simulate individual clusters with more than $10^5$
particles and with a force resolution of approximately 1--2\% of the virial
radius (Dubinski \& Carlberg 1991, Warren etal 1992, Summers etal 1996, Tormen
etal 1996, 1997, 1998).  These simulations still produced smooth dark matter
halos with very little surviving dark matter subhalos -- although interesting
results were found concerning the inner structure of halos and the formation
and accretion history of halos.

Warren etal and Carlberg were the first to resolve the inner 10\% of a halos
virial radius and claimed a power law slope $\rho(r) \propto r^{-1}$. Later
simulations by Navarro etal (1997) revealed the remarkable scaling properties
of CDM halos across a wide range of mass scales, from dwarf galaxies to galaxy
clusters. Determining the central cuspy profile of dark matter halos is an 
important but difficult computational problem and we have seen recent progress 
in this area from many groups.

Analytic work concerning the overmerging problem began with White \& Rees
(1978) who discussed the various numerical and physical processes that could
lead to the loss of substructure.  Carlberg (1994) argued that particle-halo
heating was the cause of overmerging, a result also claimed by van Kampen
(1995). However the timescale for this process was at least a Hubble time
whereas overmerging appeared on an orbital timescale. Moore etal (1996) argued
that the resolution at that time was sufficient to overcome relaxation and
tidally accelerated disruption by the finite mass background particles.

The physical processes of tidal disruption via halo-halo collisions and by the
global cluster potential take place on a more rapid timescale and could
explain the loss of substructure in the simulations prior to 1998 (Moore etal
1996 -- the investigation of the importance of this process lead to the notion
of galaxy transformation in clusters by tidal heating -- galaxy harassment.)
Tidal disruption is rapidly enhanced if halos are poorly resolved in their central
regions ($\it cf$ Figure 2). The survival of dark matter substructure depends
critically on the central density structure which in turn depends on the force
and mass resolution in a simulation. For example, if halos have resolved singular
density profiles $\rho(r)\propto r^{-2}$, then it is physically impossible to
entirely disrupt a satellite halo. Only through physical merging can such
halos be lost.

The key motivation behind the Seattle HPCC group
(http://www-hpcc.astro.washington.edu/) was to construct a high performance
parallel treecode (PKDGRAV) that could simulate the formation of cosmic
structures with high resolution. The aim was to resolve $\it sub-galactic$
halos with thousands of particles and $\it \sim kpc$ force resolution. After a
long testing program a set of accuracy parameters were chosen that allowed the
correct growth of fluctuations to be followed on large scales, as well as
accurately following the orbits of particles in the most dense regions.

\

\psfig{figure=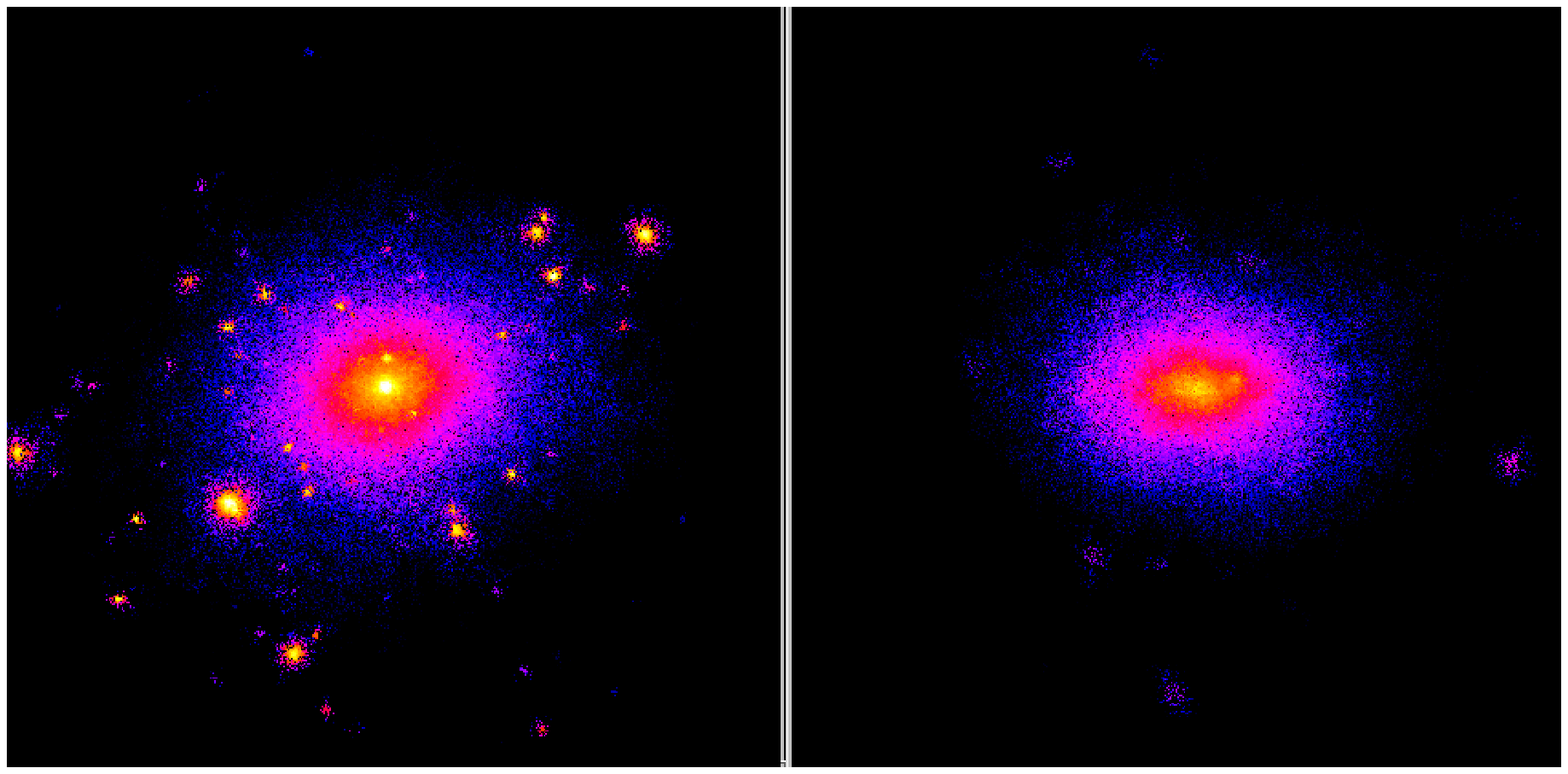,height=2.8in}
%\centerline{\epsfysize=2.55truein \epsfbox{softening.ps}}

\

\noindent{\small {\bf Figure 2} \ \ 
The same cluster simulated in Figure 1 but with two different values of the
softening length and keeping the particle mass fixed.  The left panel shows a
close up view of the inner 500 kpc of the last frame of Figure 1. In this case
the softening was 0.2\% of the virial radius. The right panel shows the same
region of the same cluster but simulated with a softening length of 1.5\% of
the virial radius.  The lack of substructure halos in the right panel
demonstrates that softened halos are easily disrupted by tidal forces.
}

\

The first simulation to finish was the 1.3 million particle ``Virgo'' cluster
simulation presented in Moore etal (1998) and analysed by Ghigna etal
(1998). This was a simulation of the formation of a single galaxy cluster
within a standard CDM universe (Figure 1). The box size was 100 Mpc but a
series of nested shells of different mass particles allowed the cluster to be
resolved at high resolution. Many hundreds of dark matter subhalos could be
found orbiting within the cluster - the overmerging problem was resolved. A
higher resolution study of the same cluster demonstrates that the central
density profile appears to have converged as have the global properties of the
subhalos within the cluster (Ghigna etal 2000).  (Similar high resolution
studies were being carried out at the same time by several other groups (c.f.
Klypin etal 1999, Colin 1999, Okamoto \& Habe 1999, Jing 2000, Fukushige \&
Makino 2000).

It is interesting to see how numerical calculations are scaling with respect
to algorithmic and computational developments.  The cluster simulation by
Peebles (1970) contained 300 particles and had a force resolution of order 100
kpc. The highest resolution simulation published to date is by Ghigna etal
(2000) in which a cluster halo was simulated with 10 million dark matter
particles and force softening of 1 kpc.  Roughly a factor of $10^7$ times the
computational cost in a timescale of 30 years -- an increase in speed that
arises from both algorithmic and hardware developments.  Four major
development have lead to this performance increase: grid based codes or
treecodes reduce the work of long range force calculations, multistepping
saves moving all the particles on the same short timestep as the handful of
particles in dense regions, faster cpu's, parallel codes can scale up to
90\% efficiency on large numbers of nodes. Figure 3 shows the ``N'' versus
year plot for several cluster simulations over the last 38 years.

\

\psfig{figure=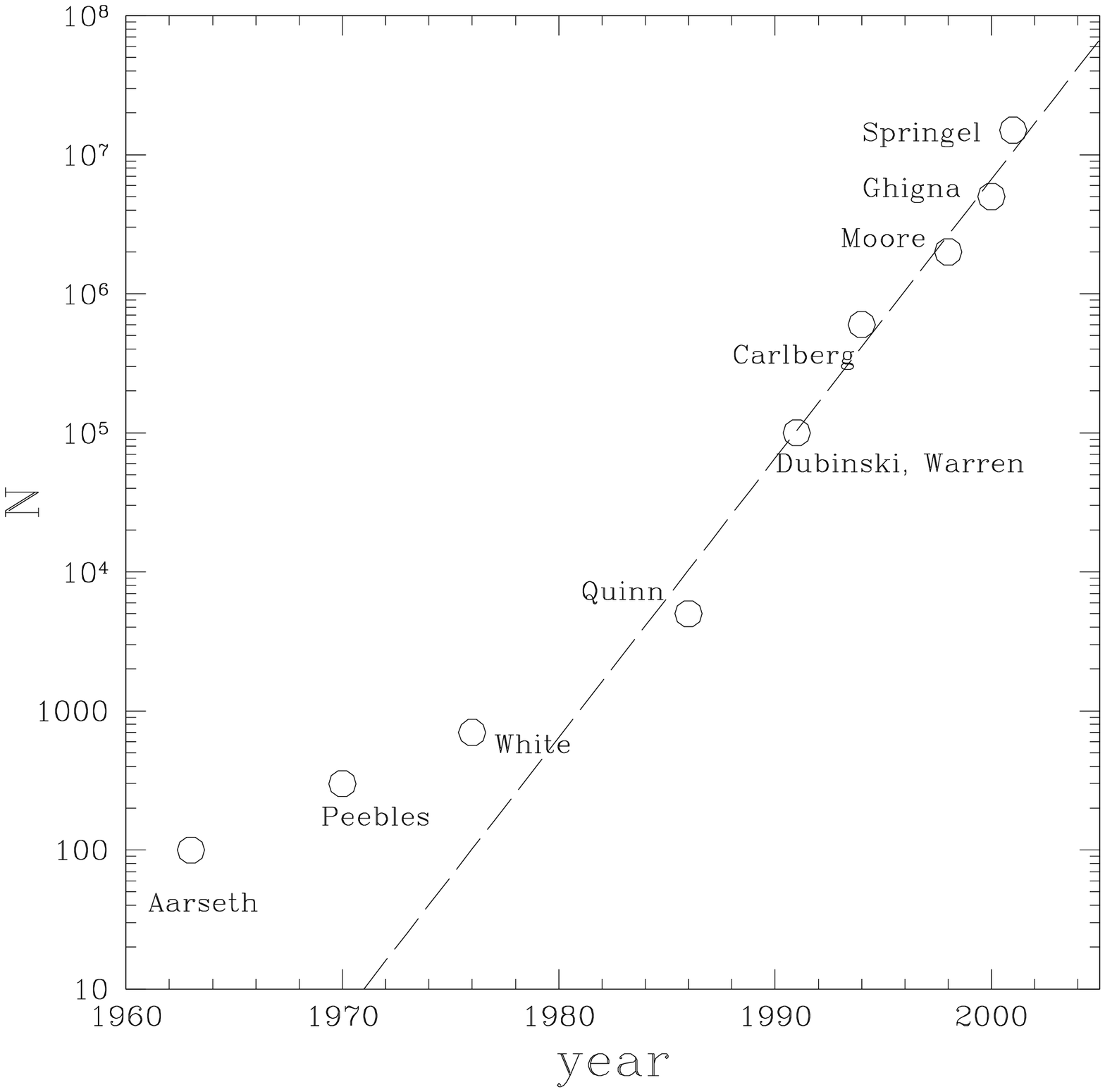,height=4.0in}
%\centerline{\epsfysize=8.0truein \epsfbox{pic1_times.ps}}

\

\noindent{\small {\bf Figure 3} \ \ 
A selection of dark matter only cluster simulations. I have attempted to plot
the number of particles within the final virial radius against the publication
year. The dashed line illustrates ``Moore's law'' (named after Gordon Moore
who founded Intel and noted the rapid increase in the density of transistors on a
silicon chip with time) which roughly equates to a doubling of cpu speed every
three years.  This type of plot suffers from many biases and is not meant to
be taken seriously. The fact that recent simulations lie on this curve does
not imply that algorithmic advances are not important -- quite the opposite
since the force softening is reduced typically in proportion to $N^{-1/3}$
which requires more computational work.
}

\

%\section{How to test against observations?}

We now have a host of new observational tests of the hierarchical structure
formation paradigm that include: the extent of galactic halos in clusters or
satellites in galactic halos, the orbital distribution of cluster/satellite
galaxies, the number of satellites as a function of their circular velocity or
mass, the spatial and velocity distribution of satellites, the central density
profiles of clusters and galaxies are close to $r^{-1.5}$.  Unfortunately,
observational tests of these model predictions on cluster scales are non
trivial and just a few exist in the literature (Tyson etal 1998, Natarajan
etal 1999, Smith etal 2000).

\end{document}